\documentclass{article}

\usepackage{microtype}
\usepackage{graphicx}
\usepackage{subfigure}
\usepackage{booktabs} 

\usepackage{amsmath}
\usepackage{textcomp}
\usepackage{amssymb}

\usepackage{hyperref}

\graphicspath{{assets/}}
\usepackage[accepted]{src/icml2020}

\icmltitlerunning{Lifelong Control of Off-grid Microgrid with Model Based Reinforcement Learning}

\newcommand{\ubar}[1]{\text{\b{$#1$}}}

\newcommand{\maxcharge}{\ensuremath{\overline{S}}}
\newcommand{\mincharge}{\ensuremath{\underline{S}}}
\newcommand{\chargerate}{\ensuremath{\overline{P}}}
\newcommand{\dischargerate}{\ensuremath{\underline{P}}}

\newcommand{\chargeEfficienty}{\ensuremath{\eta^{\text{ch}}}}
\newcommand{\dischargeEfficienty}{\ensuremath{\eta^{\text{dis}}}}

\newcommand{\charge}{\ensuremath{P^{\text{ch}}}}
\newcommand{\discharge}{\ensuremath{P^{\text{dis}}}}

\newcommand{\OPprice}[1]{\ensuremath{\pi^{\text{#1}}}}

\newcommand{\OPcost}[1]{\ensuremath{c^\text{#1}}}

\newcommand{\OPduration}{\ensuremath{\Delta_t}}

\newcommand{\curtail} {\ensuremath{P^\text{curt}}}
\newcommand{\shed} {\ensuremath{P^\text{shed}}}
\newcommand{\steer}{\ensuremath{P^\text{gen}}}
\newcommand{\nonSteerable}{\ensuremath{P^\text{res}}}
\newcommand{\steerable}{\ensuremath{P^\text{gen}}}

\newcommand{\nonFlexible}{\ensuremath{C}}

\begin{document}

\twocolumn[
\icmltitle{Lifelong Control of Off-grid Microgrid with Model Based Reinforcement Learning}



\icmlsetsymbol{equal}{*}

\begin{icmlauthorlist}
\icmlauthor{Simone Totaro}{equal,upf}
\icmlauthor{Ioannis Boukas}{equal,be}
\icmlauthor{Anders Jonsson}{upf}
\icmlauthor{Bertrand Corn\'elusse}{be}
\end{icmlauthorlist}

\icmlaffiliation{upf}{Department
of Information and Communication Technologies\\ Universitat Pompeu Fabra, Barcelona, Spain.}

\icmlaffiliation{be}{Department of Electrical Engineering and Computer Science, University of Li\`ege, Li\`ege, Belgium.} 

\icmlcorrespondingauthor{Ioannis Boukas}{ioannis.boukas@uliege.be} 

\icmlkeywords{Microgrid control, optimization, reinforcement learning, dyna.}

\vskip 0.3in
]



\printAffiliationsAndNotice{\icmlEqualContribution} 

\begin{abstract}
	The lifelong control problem of an off-grid microgrid is composed of two tasks, namely estimation of the condition of the microgrid devices and operational planning accounting for the uncertainties by forecasting the future consumption and the renewable production. The main challenge for the effective control arises from the various changes that take place over time. In this paper, we present an open-source reinforcement framework for the modeling of an off-grid microgrid for rural electrification. The lifelong control problem of an isolated microgrid is formulated as a Markov Decision Process (MDP). We categorize the set of changes that can occur in progressive and abrupt changes. We propose a novel model based reinforcement learning algorithm that is able to address both types of changes. In particular the proposed algorithm demonstrates generalisation properties, transfer capabilities and better robustness in case of fast-changing system dynamics. The proposed algorithm is compared against a rule-based policy and a model predictive controller with look-ahead. The results show that the trained agent is able to outperform both benchmarks in the lifelong setting where the system dynamics are changing over time.
\end{abstract}

\section{ Introduction} \label{sec: Introduction}
	Microgrids are small electrical networks composed of flexible consumption, distributed power generation (renewable and/or conventional) and storage devices. The operation of a microgrid is optimized in order to satisfy the demand while ensuring maximum reliability and power quality and to maximize the renewable energy harvested locally while minimizing the total system cost.

	Centralized microgrid control is usually decomposed in four tasks: i) estimating the parameters of the microgrid devices (for instance the charge efficiency of a battery storage device as a function of the state of charge and temperature, or the actual capacity of a battery after a number of cycles), ii) forecasting the consumption and the renewable production, iii) operational planning to anticipate weather effects and human activities, and iv) real-time control to adapt planned decisions to the current situation. These tasks are preformed sequentially during the lifetime of a microgrid in order to achieve near optimal operation and to maximize the benefits arising from distributed generation. 

	The estimation of the system parameters is a critical task for the optimization of the microgrid operation. The most important parameters are the operation costs as well as the capacities of the different components and the battery efficiency \cite{Parisio2014}. This process is usually carried out using measured data and is very specific to each microgrid configuration. These parameters are then used in the simulation model where the system operation is modeled.

	After the parameters' estimation, it is important for the efficient microgrid operation to incorporate in the decision making process all the sources of uncertainty. To this end, forecasting techniques are deployed for the stochastic production and consumption. Forecasts are collected several times before the physical delivery in order to improve the accuracy of the forecasted value in the light of new information. There is a variety of forecasting techniques in the literature ranging from fundamental models of consumption and renewable energy production \cite{Dolara2015} to statistical models using measured data \cite{Lombardi2019}.

	Subsequently, the outputs of the forecasting models in combination with the system parameters are used to compute the optimal control actions that need to be taken. The optimization of the control actions can be performed using the simulation model of the microgrid. However, the nonlinearities introduced by the system components make this problem hard to solve and without any optimality guarantees. Therefore, it is common in the literature to use a mixed integer linear approximation of the system model that can be solved easily with modern techniques and to optimality. A rolling horizon strategy is then usually adopted where the optimization is performed with some predefined look-ahead period \cite{Palma-Behnke2013}. Alternatively, a model predictive control (MPC) strategy is used for achieving economic efficiency in microgrid operation management \cite{Parisio2014}. An MPC policy is a feedback control law meant to compensate for the realization of uncertainty.

	Given the data availability, the two preceding tasks, namely forecasting and optimization, can be merged into one task and a control action can be derived directly from the data observed. To this end, reinforcement learning can be leveraged as a methodology to deal with the uncertainty and the non-linearities of the system components. The benefit from this approach is twofold: i) the non-linearities of the simulator are accounted for in the optimization process; and ii) the training of parameters is performed in the direction of an objective function that corresponds to the system cost instead of the mean squared error as in the case of forecasting methods.

	In this paper, we present an open-source reinforcement framework for the modeling of an off-grid microgrid for rural electrification. Moreover, we formulate the control problem of an isolated microgrid as a Markov Decision Process (MDP). The degradation of the various components as well as the non-linear dynamics of several devices are considered. Due to the high-dimensional continuous action space we define a set of discrete meta-actions in a similar way to previous work~\cite{Boukas2018}.

	The main challenge for the lifelong control of an off-grid microgrid arises from the uncertainty of the future renewable production and consumption. A critical issue in microgrid operation is that oftentimes the policy learned during training on a dataset does not perform well on unseen data. Additionally, the degradation or damage of the various components such as the storage devices or the photovoltaic panels cause the previously learned policies to become sub-optimal over time.

	To address these challenges we propose a novel model based algorithm. In particular, the algorithm is an instance of \textsc{Dyna} \cite{sutton2012dyna}, where the model is trained using distributional losses and the policy is optimized using the Proximal Policy Optimization (PPO) algorithm. The values of the policy are updated based on the expectation computed over a set of states sampled from a model. This model is trained online using samples from the real environment. We illustrate that this algorithm allows for much better estimation of the values accounting for the uncertainties and yields enhanced exploration. Additionally, we show that the enhanced exploration gained using the model to sample states allows for better generalization to unseen data. Moreover, we show that the knowledge of a previously trained policy can be efficiently transferred when training on a new set of data. Finally, we demonstrate the ability of the controller to adapt to sudden changes such as damage of the equipment without explicit knowledge of the event. 

	To evaluate the performance of the obtained policy, we compare it with two benchmarks: i) a rule-based control that takes decisions in a myopic manner based only on current information; and ii) an optimization-based controller in which look-ahead is applied to forecast consumption and production.

	This paper is organized as follows. Section~\ref{sec: RelatedWork} elaborates on state-of-the-art methods used for microgrid operation and control. Section~\ref{sec: RLBackground} provides the theoretical background used for the developed framework and the algorithm proposed. In Section~\ref{sec: Simulator}, the system dynamics of the microgrid are detailed. In Section~\ref{sec: ProblemStatement}, we formulate the lifelong control problem of an off-grid microgrid as an MDP. Section~\ref{sec: Algorithm} presents the model based algorithm used to solve the lifelong microgrid control problem. The proposed algorithm is compared against the two benchmark strategies presented in Section~\ref{sec: Bechmarks}. Section~\ref{sec: CaseStudy} describes the case study and results obtained. Finally, Section~\ref{sec: conclusions} concludes the main findings and provides avenues for future research.
\section{Related Work} \label{sec: RelatedWork}

	The control of a system that includes various sources of uncertainty like the case of an off-grid microgrid heavily depends on the availability of accurate values of forecasts of the variability. Following the recent advances in artificial intelligence and the data availability, the forecasting of renewable energy sources (RES) using artificial neural networks (ANNs) has been proposed~\cite{Leva2019}, where a hybrid model combines ANNs with a Clear Sky Solar Radiation Model (CSRM) for the solar power forecasting. The resulting model is using weather forecasts and real hourly photovoltaic power data. Short-term load forecasting (STLF) for microgrids using artificial intelligence has also been proposed~\cite{Hernandez2014}, using a three-stage architecture in which a self-organizing map (SOM) is applied to the input. The outputs are clustered using the k-means algorithm, and finally demand forecasting for each cluster is performed using an ANN.

	Apart from the conventional control approaches of using a simplified model of the system and some forecasts of the variable sources to optimize the operation of the microgrid, some data-driven approaches have been proposed in the literature. One example is a modeling framework for the control of the storage device in the context of an interconnected microgrid~\cite{Kuznetsova2013}. Optimal Q-values are computed using the Q-learning method. In this setting the state and action spaces are discretized in order to reduce the computational complexity and the results show increased utilization of RES production compared to optimization methods. In the present paper, we use a more detailed representation of the system to enable more complex and expressive policies.

	Following the recent advancements in the field of Deep Reinforcement Learning (DRL), researchers have also proposed a Deep Q-learning approach for the control of seasonal storage in an isolated microgrid~\cite{franccois2016deep}. In this framework, a specific deep learning structure is presented in order to extract information from the past RES production and consumption as well as the available forecasts. Despite the highly dimensional continuous state space, the authors obtain a control policy that is able to utilize the long-term storage in a meaningful way. However, in this approach it is assumed that the dynamics of the system are linear and that forecasts of the variable resources are available.

\section{Reinforcement Learning Background} \label{sec: RLBackground}

\subsection{ Markov Decision Process}

    We consider an infinite horizon discounted Markov Decision Process (MDP), defined by the tuple $\langle S,A,r,\{P_t\}_t,\gamma \rangle$ where  $S$ is the state space, $A$ the action space, $r:S \times A \rightarrow \mathbb{R}$ is the Markovian cost function, $P_t: S \times A \rightarrow \Delta(S) $, $t\geq 0$, is the transition kernel at time $t$ and $\gamma \in (0, 1) $ is the discount factor. Here, $\Delta(S)$ is the probability simplex on $S$, i.e.~the set of all probability distributions over $S$. At each time step $t$, the agent observes state $s_t \in S$, takes an action $a_t \in A$, obtains reward $r_t$ with expected value $\mathbb{E}[r_t] = r(s_t, a_t)$, and transitions to a new state $s_{t+1} \sim P_t(\cdot \rvert s_t, a_t)$. We refer to $(s_t,a_t,r_t,s_{t+1})$ as a {\em transition}. Note that the transition kernels may not be stationary.
   
    Let $\pi$ denote a stochastic policy $\pi : S \rightarrow \Delta(A)$ and $\eta(\pi)$ its expected discounted cumulative reward under some initial distribution $d_0\in\Delta(S)$ over states: 
    \begin{gather}
    \eta(\pi) = E_{s\sim d_0} [V^{\pi}(s)],\label{eqn: expected_cumulative_reward}
    \end{gather}
    
	\noindent
     where $\tau = \{(s_t,a_t, r_t)\}_{t \geq 0}$ is a trajectory, $p(\tau)$ is the probability distribution over trajectories,  
     \begin{equation}
        p(\tau) = d_0(s_0) \prod_{t=0}^\infty P_t(s_{t+1} \rvert s_t, a_t) \pi(a_t \rvert s_t),
     \end{equation}{}
     
 	\noindent
    and the value function $V^\pi$ is defined for each state $s\in S$ as
    \begin{gather}
        V^{\pi}(s) = E_{p(\tau)}\left[ \left. \sum_{t=0}^{\infty} \gamma^t r_t(s_t,a_t) \right\vert s_0=s \right].
    \end{gather}

    The goal of the agent is to find a policy that maximizes the expected cumulative reward $\eta(\pi)$:
    \begin{gather}
    \eta^{*} = \max_{\pi} \eta(\pi),\label{eqn: optimalcost}\\
    \pi^{*} = \arg \max_{\pi} \eta(\pi).\label{eqn: optimalpolicy}
    \end{gather}

    \begin{algorithm}[t]
    	\caption{\textsc{Dyna}}
    	\begin{algorithmic}[1]
    		\STATE \textbf{Inputs}: MDP $M$, integers $T$, $B$, $N$
			\STATE initialize policy $\pi_\theta$, model $M_\psi$
    		\FOR{$t=0$ \TO $T-1$}
    			\STATE $s \sim d_0$
    			\STATE $a \sim \pi_\theta(\cdot \rvert s)$
    			\STATE $s',r \sim M(s,a)$
				\STATE $\pi_\theta = \,$ \textsc{updatePolicy}$(s,a,r, s')$
    			\STATE $M_\psi = \,$ \textsc{updateModel}$(s,a,r, s')$
    			\IF{$t \geq B$}
        			\FOR{$n=0$ \TO $N-1$}
        				\STATE  $s \sim d_0$
        				\STATE  $a \sim \pi_\theta(\cdot \rvert s)$
        				\STATE $\hat{s}', \hat{r} \sim M_\psi(s,a)$
        				\STATE $\pi_\theta = \,$ \textsc{updatePolicy}$(s,a,\hat{r}, \hat{s}')$
        			\ENDFOR
        		\ENDIF
    		\ENDFOR
    	\end{algorithmic}
    	\label{algodyna}
    \end{algorithm}
    
\subsection{ Dyna}
    \textsc{Dyna} \cite{sutton2012dyna} is a model based reinforcement learning architecture that aims to integrate learning and planning. 
    It does so by performing online estimation of the transition kernel and reward function. Let $M_\psi = \langle P_\psi, r_\psi \rangle$ be a parametric model learned during training. Note that we estimate a single transition kernel $P_\psi$ even though the true kernel may not be stationary.
    
	Algorithm~\ref{algodyna} outlines the \textsc{Dyna} algorithm in the parametric setting. 
    For every transition $(s,a,r,s')$ sampled from the environment $M$, we update the policy $\pi_\theta$ and parametric model $M_\psi$ via update functions described in Algorithms~\ref{update_policy} and \ref{update_model}. We remark that the policy update typically relies on a value function $V_\phi$, and that the value function and the components of the parametric model are updated by minimizing a loss function.
 
    After the update step, we use the learned model to perform $N$ updates of the policy $\pi_\theta$, in the same way as one would using the true environment. At every step we sample a state $s \sim d_0$, apply action $a \sim \pi_\theta( \cdot \rvert s) $ and query the parametric model $\hat{s}', \hat{r} \sim M_\psi(s,a)$. 
    
    Note that there are two main differences during the planning phase. First, the transition $(s, a, \hat{r}, \hat{s}')$ comes from the parametric model, and second, there is no structure in the sampling process, therefore in such an update the agent can experience any possible one step transition, even ones that are hard to gather under the current policy.

\begin{algorithm}[t]
	\caption{\textsc{updatePolicy}}
	\begin{algorithmic}[1]
		\STATE \textbf{Input}: transition $(s,a,r,s')$
		\STATE  $V_\phi = \arg\min_{V_\varphi} L^V(V_\varphi)$
		\STATE  $\pi_\theta = \arg\max_{\pi_\varphi} \eta (\pi_\varphi)$
	\end{algorithmic}
	\label{update_policy}
\end{algorithm}

\begin{algorithm}[t]
	\caption{\textsc{updateModel}}
	\begin{algorithmic}[1]
		\STATE \textbf{Input}: transition $(s,a,r,s')$
		\STATE  $P_\psi = \arg\min_{P_\varphi} L^P(P_\varphi)$
		\STATE  $r_\psi = \arg\min_{r_\varphi} L^r(r_\varphi)$
	\end{algorithmic}
	\label{update_model}
\end{algorithm}

\subsection{ Proximal Policy Optimization}
    
    The Proximal Policy Optimization (PPO) algorithm \cite{Schulman2017} belongs to the family of policy gradient methods and can be used with both discrete and continuous action spaces. In the vanilla actor-critic method \cite{sutton2000policy}, a stochastic policy $\pi_\theta$ with parameters $\theta$ is optimized towards the following regularized objective:
    
    \begin{gather}
        \eta(\pi_\theta)  = \mathbb{E}_{p(\tau)}\left\lbrace 
        \frac{\pi_\theta(a_t \rvert s_t)}{\pi_o(a_t \rvert s_t  )} \hat{A}_{\phi}(s_t,a_t)\right\rbrace  - \frac{1}{\beta}  
        D( \pi_\theta \rvert \rvert \pi_o) ,\label{eqn: regularized}
    \end{gather}

	\noindent
	where $\pi_o$ is the old policy, $\hat{A}_{\phi}(s_t,a_t)$ is an estimator of the advantage function, $D$ is a regularizer in the form of a Bregman divergence and $\beta$ is a learning rate.
    
	Since \eqref{eqn: regularized} is hard to optimize directly, the policy is repeatedly updated using stochastic gradient descent. Concretely, a gradient step is used to update of the parameters $\theta$ as
    \begin{gather}
    \theta_{new} = \theta + \alpha \nabla \hat{\eta}(\pi_{\theta}),\label{eqn: thetaupdatesgrad}
    \end{gather}
    
	\noindent
    where $\alpha$ is a step size and the regularized objective for the individual transition $(s,a,r,s')$ is estimated as
	\begin{gather}
        \hat{\eta}(\pi_\theta)  =  \frac{\pi_\theta(a \rvert s)}{\pi_o(a \rvert s )} \hat{A}_{\phi}(s,a)  - \frac{1}{\beta}  
        \sum_{a'} \pi_\theta(a'|s) \log \frac {\pi_\theta(a'|s)} {\pi_o(a'|s)}. \label{eqn: regularized2}
    \end{gather}

	An unbiased estimator of the advantage function is given by 
    \begin{equation}
    \hat{A}_{\phi}(s,a) =  r + \gamma \hat{V}_{\phi} (s') - \hat{V}_{\phi}(s), \label{eqn: advantage}
    \end{equation}
    
	\noindent
    where the estimated value function $\hat{V}_{\phi}$ is obtained by minimizing the following loss:
    \begin{equation}
        L^V(\hat{V}_{\phi}) = \frac{1}{2} \mathbb{E}_{p(\tau)} [ \hat{A}_{\phi_{old}}(s_t, a_t)^2].
    \end{equation}

    In practice, rather than performing updates for individual transitions, the algorithm performs multiple epochs of mini-batch updates of stochastic gradient descent of both the policy and value function.

\subsection{Quantile Regression}


	The problem of estimating a model $M_\psi = \langle P_\psi, r_\psi \rangle$ is commonly cast as supervised learning, in which the components of $M_\psi$ are computed by minimizing loss functions. One of the contributions of our proposed algorithm is to use distributional losses to estimate $M_\psi$ in the parametric setting.

	Distributional losses introduced by \citet{bellemare2017distributional} and expanded by \citet{dabney2018distributional} achieve state of the art performance in several reinforcement learning benchmarks. \citet{imani2018improving} discuss the importance of distributional losses for regression problems, arguing that such losses have locally stable gradients which improves generalization. Here we concisely describe the loss function that we use in our setting. For a more detailed description the reader can consult \citet{dabney2018distributional}. 

	Our goal is to learn the distribution of some random variable $z \sim F(z)$. To do so, it is known that the value of the quantile function $F^{-1}_z(\tau)$ is the minimizer of the quantile regression loss. Let $(-k, k)$ be the support of the empirical CDF. The Quantile Huber loss is defined as
		\begin{equation}
			\rho_{\tau}(u) = \lvert \tau - \delta_{\{u\leq0\}}\rvert L(u),
		\end{equation}
	where $L(u)$ is given by
		\begin{equation}
			L(u) = 
			\begin{cases}
				\frac{1}{2} u^2,& \text{if} \lvert u \rvert \leq k, \\
				k(\lvert u\rvert - \frac{1}{2}k),& \text{otherwise}.
			\end{cases}
		\end{equation}
In Section~\ref{sec: Algorithm} we show how to adapt this loss to learn the estimated transition kernel $P_\psi$ and reward $r_\psi$.

\section{Microgrid Description}\label{sec: Simulator}
    \label{sec:gym}
    In this section we provide a detailed description of the system considered. An off-grid microgrid designed for rural electrification is inherently characterized by changes occurring in different time-scales. We provide a formal description of the different types of changes and we motivate the need for a lifelong control that has the ability to adapt to these changes.

\subsection{Components}

An off-grid microgrid is composed of the following components:
 
\subsubsection{Consumption} 
    
    The consumption of the isolated microgrid $\nonFlexible$ is considered to be non-flexible, meaning that there is a high cost associated to the energy non-served. The consumption $\nonFlexible_{t}$ at each time-step $t$ of the simulation is assumed to be a stochastic variable that is sampled from distribution $P^{C}_{t}$, given the $h$ previous realizations, according to:
    \begin{gather}
    \nonFlexible_{t} \sim P^{C}_{t}( \nonFlexible_{t-1}, ...,\nonFlexible_{t-h} ).\label{eqn: load}
    \end{gather}
    In this paper, it is represented by real data gathered from an off-grid microgrid. The distribution $P^{C}_{t}$ is indexed in time in order to indicate that changes occur in the aggregate consumption over the life-time of the microgrid. For instance, a change in the consumption profile can be caused by the fact that more users are progressively connected to the micro-grid.

\subsubsection{Storage model}
    
    The modeling of the storage system can become quite complex and highly-nonlinear depending on the degree of accuracy required by each specific application. In this paper, we use a linear ``tank'' model for the simulation of the battery since we assume that the simulation time-step size $\Delta t$ is large enough (1 hour). The dynamics of a battery are given by:
    \begin{gather}
    SoC_{t+1} = SoC_{t}+ \Delta t\cdot(\chargeEfficienty \charge_t - \frac{\discharge_t}{\dischargeEfficienty}) ,\label{eqn: storagedynamics}
    \end{gather}
    where $SoC_{t}$ denotes the state of charge at each time step $t$, $\charge$ and $\discharge$ correspond to the charging and discharging power, respectively and $\chargeEfficienty$, $\dischargeEfficienty$ represent the charging and discharging efficiencies of the storage system. The charging ($\charge$) and discharging ($\discharge$) power of the battery are assumed to be limited by a maximum charging rate $\chargerate$ and discharging rate $\dischargerate$, respectively. Accounting for the storage system degradation, we consider that the maximum capacity $\maxcharge$ of the storage system as well as the charging and discharging efficiencies ($\chargeEfficienty$, $\dischargeEfficienty$) are decreasing as a linear function of the number of cycles $n_t$ that are performed at each time-step $t$. We have, $\forall t \in T$, 
    \begin{align}
    SoC_{t}, \charge_t, \discharge_t &\geq 0 \\
    \charge_t &\leq \chargerate\\
    \discharge_t &\leq \dischargerate , \\
    SoC_{t} &\leq \maxcharge \label{eqn: storage limits},\\
    \maxcharge&=s(n_t). \label{eqn: storage_limits_evolving}
    \end{align}

\subsubsection{Steerable generator model}
    Steerable generation is considered any type of conventional fossil-fuel based generation that can be dispatched at any time-step $t$. When a generator is activated, it is assumed to operate at the output level $\steer_t$ that is ranging between the minimum stable generation $\underline{\steerable}$ and the maximum capacity $\overline{\steerable}$ such that:
    \begin{gather}
    \underline{\steerable} \leq \steer_t \leq \overline{\steerable}.\label{eqn: generatordynamics}
    \end{gather}
    The fuel consumption $F_t$ related to the operation of the generator at time $t$ is a linear function of the power output $\steer_t$ curve with parameters $F_1$, $F_2$ given by the manufacturer. 
    \begin{gather}
    F_t = F_1 + F_2 \cdot \steer_t .\label{eqn: fuelconsuption}
    \end{gather}
    The fuel cost $\OPcost{fuel}_t$ accounting for the fuel price $\OPprice{steer}$ is then given by:
    \begin{gather}
    \OPcost{fuel}_t = F_t \cdot \OPprice{fuel} .\label{eqn: generatorcost}
    \end{gather}
    
\subsubsection{Non-steerable generators model}
    The level of non-steerable generation from renewable resources such as wind or solar is denoted by $\nonSteerable$. Similar to the non-flexible load case it is assumed that $\nonSteerable_t$ at time-step $t$ is sampled from a probability distribution $P^{\nonSteerable}_{t}$, given the $h$ previous realizations, according to:
    \begin{gather}
    \nonSteerable_t \sim P^{\nonSteerable}_{t}( \nonSteerable_{t-1},...,\nonSteerable_{t-h}).\label{eqn: pv}
    \end{gather}
    In this paper, the renewable generation is represented by real data gathered from an off-grid microgrid. Similar to the case of the non-flexible load, the distribution $P^{\nonSteerable}_{t}$ is indexed by time $t$ to indicate that changes in the renewable production might occur over time. These changes are mostly related to the progressive degradation of the equipment (solar panels).
    
\subsubsection{Power balance}
    At each time-step $t$ in the simulation horizon we compute the power balance between the injections and the off-takes. The residual power resulting from the mismatch between production and consumption is curtailed $\curtail_t$ if its positive and shed $\shed_t$ if it is negative. We can formally define the power balance as:
    \begin{gather}
    \nonSteerable_{t} + \steer_{t} + \discharge_{t} + \shed_{t} \\\notag
    =\charge_{t} + \curtail_{t}  + \nonFlexible_{t},\label{eqn: powerbalance}
    \end{gather}
    with $\curtail_t , \shed_t \geq 0$.
    The costs arising from the curtailment of generation or the shedding of non-flexible loads are given by: 
    \begin{gather}
    \OPcost{curt}_t = \curtail_t \cdot \OPprice{curt} \label{eqn: curtcost}\\
    \OPcost{shed}_t = \shed_t \cdot \OPprice{shed} \label{eqn: shedcost}
    \end{gather}

\subsection{Characterizing changes in the environment}
	
	Oftentimes in real-life applications the concept of interest depends on some underlying context that is not fully observable. Changes in this underlying concept might induce more or less radical changes in the concept of interest, which is formally known as concept drift \cite{tsymbal2004problem}. For instance, in the off-grid microgrid under study the connection of new users and their habits have strong influence on distribution $ P^{C}_{t}$. However, it is not possible to know exactly and to quantify the effect on the consumption a priori. 
	

	In this paper, we deal with the following two distinct set of changes: 1) gradual changes that affect the non-controllable dynamics; and 2) sudden changes that affect the deterministic dynamics.
	As described in Section~\ref{sec: ProblemStatement}, one can decouple the two components of the state space. Gradual changes occurs in the stochastic component of the state space (\ref{eqn: stochastic_probabilities}) while sudden changes occurs in the deterministic system dynamics (\ref{eqn: transitionfunction}).


	\subsubsection{Gradual changes}
	These are cases in which a slow concept drift occurs. The extent of the drift is bounded so that any learner can follow these changes successfully. A formal bound on the maximal rate of drift that is acceptable by a batch-based learner is given by \citet{kuh1991learning}.
	
	In this paper we assume that changes related to the consumption and renewable production profiles as well as degradation of the equipment (storage) belong to this category.
	
	\subsubsection{Sudden or abrupt changes}
	In our setting, sudden or abrupt changes are adversarial changes that affect the system dynamics, and for which the learner needs to find the best response. Robust MDPs \cite{nilim2005robust} describes optimal control under such changes and recent work \cite{lim2013reinforcement} shows that incorporating learning in such contexts can deliver policies as good as the minimax policy.
	\citet{gajane2018ucrl} also propose an algorithm for detecting abrupt changes in MDPs.
	In the concept of an off-grid micro-grid this type of change would typically occur during equipment failure.

\section{Problem Statement} \label{sec: ProblemStatement}

    The operation of the system described in section \ref{sec: Simulator} can be modelled as a Markov Decision process as it is defined in Section \ref{sec: RLBackground}. We consider that at each time-step $t \in T$ the state variable $s_t \in S$ is composed of a deterministic and a stochastic part as $s_t = \left(\ubar{s}_t, \bar{s}_t\right) \in S$ and contains all the relevant information for the optimization of the system. The deterministic part $\ubar{s}_t = \left(SoC_{t}\right) \in \ubar{S}$ corresponds to the evolution of the state of charge of the storage device and can be fully determined by Equations (\ref{eqn: storagedynamics})-(\ref{eqn: storage_limits_evolving}). The stochastic variable $\bar{s}_t$ represents the variable renewable production and consumption as $\bar{s}_t = \left(\left(\nonFlexible_t,...,\nonFlexible_{t-h}\right), \left(\nonSteerable_t,...,\nonSteerable_{t-h} \right)\right) \in \bar{S}$ as defined in Equations (\ref{eqn: load}) and (\ref{eqn: pv}).
    
    The available control action $a_t$ that can be applied at each time-step $t$ is defined as:
    \begin{gather}
    a_t= \left(\charge_{t}, \discharge_{t}, \steer_{t} \right) \in A,
    \end{gather}
    and contains the charging/discharging decision for the storage system and the generation level of the steerable generators.
    
    At each time-step $t$ the system performs a transition based on the dynamics described in Section~\ref{sec: Simulator} according to
    \begin{gather}
    \ubar{s}_{t+1} = f_{t}\left(s_t, a_t\right), \label{eqn: transitionfunction}\\
    \bar{s}_{t+1} \sim \bar{P}_{t}\left(\bar{s}_t\right)\label{eqn: stochastic_probabilities},
    \end{gather}
    where $f_{t}$ is a deterministic function and $\bar{P}_{t}$ is used to denote the joint probability distribution of the stochastic variables $\nonFlexible, \nonSteerable$ as defined in Equations (\ref{eqn: load}) and (\ref{eqn: pv}). Note that, the transition function $f_{t}$ is indexed in time to account for the changes (e.g. degradation) that may occur to the equipment. Equations (\ref{eqn: transitionfunction}) and (\ref{eqn: stochastic_probabilities}) can fully determine the transition kernel of the MDP at each time step as $P_t: S \times A \rightarrow \Delta(S)$.
    
    Each transition generates a non-positive reward signal (cost) signal $r_t$ according to the reward function $r(s_t, a_t) \in \mathbb{R}$, defined as:
    \begin{gather}
    r_t = r(s_t, a_t) = -(\OPcost{fuel}_t + \OPcost{curt}_t +  \OPcost{shed}_t).\label{eqn: rewardfunction}
    \end{gather}
    
     The problem of life-long control of an off-grid microgrid is equivalent to finding a policy $\pi$ that maximizes the total expected discounted cumulative reward $\eta(\pi)$ as defined in Equations (\ref{eqn: expected_cumulative_reward})-(\ref{eqn: optimalpolicy}).

    \subsection{Microgrid Simulator}
    The described MDP for off-grid microgrid control is available as an open source simulator\footnote{Available at \url{https://github.com/bcornelusse/microgridRLsimulator}.}implemented in OpenAI gym \cite{brockman2016openai}. The simulator contains a detailed modelling of the microgrid components and allows for applying any control strategy. It receives as input the microgrid configuration (components size and parameters, time series representing the exogenous information, and simulation parameters) and simulates the operation for a predefined simulation horizon $T$.

\section{Algorithm} \label{sec: Algorithm}
        
        Real world applications are non-stationary, partially observable and high dimensional. A desirable algorithm should effectively deal with those challenges as well as provide basic safety guarantees \cite{dulac2019challenges}.
        
        Model Based RL algorithms are appealing for real world applications because they are sample efficient, they explicitly approximate the environment dynamics, and, when combined with powerful function approximation, they can scale to the high dimensional setting \cite{nagabandi2018neural}.
        
        The key issue with Model Based RL is learning the model sufficiently well to be useful for policy iteration. For real world application, this issue is worsened by the requirements of generalisation and sample efficiency.
        
        To address those challenges we propose a practical algorithm that builds upon the \textsc{Dyna} algorithm \cite{sutton2012dyna}.
        We use a variant of PPO \cite{Schulman2017} to perform policy iteration, and quantile losses to approximate the model dynamics. We have two quantile losses, one for learning the transition kernel $P_\psi$ and one for the learning the reward function $r_\psi$:
       
        \begin{equation}
        	L^P(s) = \mathbb{E}[\sum_{i=1}^q \rho_{\tau_i}(s' - P_\psi(s,a)] 
        \end{equation}
        \begin{equation}
        	L^r(r) = \mathbb{E}[\sum_{i=1}^q \rho_{\tau_i}(r - r_\psi(s,a)]
        \end{equation}
        
        Model free updates are performed in PPO by sampling a partial trajectory and directly maximing {\eqref{update_policy}}. 
        We use two seperate networks for the value $V_\phi$ and the policy $\pi_\theta$, and we select the advantage estimator as in~\eqref{eqn: advantage}. Model based updates are one-step simulated transitions. As noted in previous work~\cite{van2019use}, updating simulated states helps to empirically mitigate model error, constraining it to simulated states. Complementary work~\cite{janner2019trust} shows that simulating one-step transitions provides a strong baseline with respect to partial or complete policy rollouts with a learned model, and PPO mantains its monotonic improvement property.
        
        In practice, in order to deal with the high dimensionality of the state and action space, we represent the model $M_{\psi}$ as a neural network with shared parameters $\psi \in \Psi$ and two heads $P_\psi$ and $r_\psi$. Each head outputs a vector of size $d \times q$ where $d$ is the output dimension and $q$ is the number of quantiles considered.
        The policy $\pi_\theta$ and the value function $V_\phi$ are represented using two different networks. Contrary to previous claims~\cite{Schulman2017}, sharing parameters does not improve learning in our experiments.
        Finally, we introduce a hyperparameter $B \in \mathbb{N}$ that is the minimum amount of optimisation performed with the model prior to allowing model based updates. Empirically we found this to reduce the detrimental effect of model error on policy updates. We refer to the presented algorithm as \textsc{D-Dyna}.
        
   \section{Benchmark strategies} \label{sec: Bechmarks}

In this section, we introduce two control strategies used for comparison purposes. First, a myopic rule-based strategy is used to provide a lower bound of the total rewards in the period considered. The second strategy corresponds to a model-predictive control (MPC) with $N$-step look-ahead. In the case that we consider a sufficiently large number of look-ahead steps and full knowledge about the future realization of the stochastic variables, the MPC can provide an upper bound on the total rewards that can be obtained by a policy.

\subsection{Rule-based controller}
The rule-based controller is a simple myopic controller that implements a set of decision rules to determine the control actions that need to be taken at each time-step $t$. It requires only data regarding the present condition of the microgrid. The logic that is implemented is the following:
\begin{enumerate}
	\item First, the residual generation $\Delta P_t$ is computed as the difference between the current total renewable production and non-flexible demand as: $$\Delta P_t = \nonSteerable_t - \nonFlexible_{t}$$
	\item If $\Delta P_t$ is positive, the status of the battery is set to charge ($``C"$) and the decision $y_t$ is formed as:$$y_t = ``C"$$
	\item If $\Delta P_t$ is negative, the status of the battery is set to discharge ($``D"$) and the decision $y_t$ is formed as:$$y_t = ``D"$$
	\item When the decision $y_t$ is made, the residual generation is dispatched over devices as presented in Algorithm \ref{algoRBC}, and the control action $a_t= \left(\charge_{t}, \discharge_{t}, \steer_{t} \right)$ containing to the storage device ($\charge_{t}$, $\discharge_{t}$) and the generator ($\steer_{t}$) is determined. 
	
\end{enumerate}

\begin{algorithm}[t]
	\caption{Power dispatch.}
	\begin{algorithmic}[1]
		\STATE \textbf{Inputs:} $\Delta P_t$ , $y_t$, $\chargerate$, $\dischargerate$, $\overline{\steerable}$
		\STATE \textbf{Initialize:}$\discharge_t \gets 0$,$\charge_t\gets 0$,  $\steer_t\gets 0$
		\IF{ $\Delta P_t \geq 0 $}
		\IF{ $y_t = ``C" $}
		\STATE $\charge_t = \min(P^{RES},\chargerate)$ 
		\ENDIF
		\STATE $\Delta P_t \leftarrow \Delta P_t - \charge_t$
		\ELSE
		\IF{ $y_t = ``D" $}
		\STATE $\discharge_t = \min(-P^{RES},\dischargerate)$ 
		\STATE $\Delta P_t \leftarrow \Delta P_t +\discharge_t$
		\STATE $\steer_t=\min(-P^{RES},\overline{\steerable})$ 
		\ENDIF
		\IF{ $y_t = ``G" $}
		\STATE $\steer_t=\min(-P^{RES},\overline{\steerable})$ 
		\STATE $\Delta P_t \leftarrow \Delta P_t + \steer_t$
		\STATE $\discharge_t = \min(-P^{RES},\dischargerate)$ 
		\ENDIF
		\ENDIF
	\STATE \textbf{Output:} $a_t$
	\end{algorithmic}
	\label{algoRBC}
\end{algorithm}

\subsection{Model-predictive controller}\label{sec: optcontroller}

\begin{algorithm}[t]
	\caption{Model-predictive controller.}
	\begin{algorithmic}[1]
		\STATE \textbf{Inputs:} $N$, $\OPprice{curt}$ , $\OPprice{shed}$, $\OPprice{fuel}$,$F_1$, $F_2$, $\chargeEfficienty$, $\dischargeEfficienty$, \\
		$\chargerate$, $\dischargerate$, $\maxcharge$, $\overline{\steerable}$ , $\underline{\steerable}$, $\widehat{\nonFlexible}_{t}$,$\widehat{\nonSteerable}_{t}$
		\STATE \textbf{Solve:}
		\begin{align}
	\hspace{-12pt}	\min &  \sum_{k=0}^{N} \OPduration \big(\OPcost{fuel}_t + \OPcost{curt}_t + \OPcost{shed}_t \big)&&& \nonumber\\
	\hspace{-12pt}s.t.&\widehat{\nonSteerable}_{t+k} + \steer_{t+k} + \discharge_{t+k} + \shed_{t+k}=&&\nonumber\\
        &\hspace{0pt}\charge_{t+k} + \curtail_{t+k}  + \widehat{\nonFlexible}_{t+k} \hspace{28pt}, \forall k \in \{0,...,N-1 \}&&\nonumber\\
        &\OPcost{curt}_{t+k} = \curtail_{t+k} \cdot \OPprice{curt} \hspace{38pt}, \forall k \in \{0,...,N-1 \} &&\nonumber\\
        &\OPcost{shed}_{t+k} = \shed_{t+k} \cdot \OPprice{shed} \hspace{35pt}, \forall k \in \{0,...,N-1 \}&&\nonumber\\
        &\OPcost{fuel}_{t+k} = F_{t+k} \cdot \OPprice{fuel} \hspace{38pt}, \forall k \in \{0,...,N-1 \}&&\nonumber\\
        &F_{t+k} = F_1 + F_2 \cdot \steer_{t+k} \hspace{20pt}, \forall k \in \{0,...,N-1 \}&&\nonumber\\
        &SoC_{t+k+1} = SoC_{t+k}+ \Delta t  \cdot(\chargeEfficienty \charge_{t+k} -&&\nonumber\\
        &\hspace{50pt} \frac{\discharge_{t+k}}{\dischargeEfficienty}) \hspace{40pt}, \forall k \in \{0,...,N-1 \}&\nonumber\\
        &SoC_{t+k}, \charge_{t+k}, \discharge_{t+k} \geq 0 \hspace{5pt}, \forall k \in \{0,...,N-1 \}&&\nonumber\\
        &\charge_{t+k} \leq \chargerate \hspace{75pt}, \forall k \in \{0,...,N-1 \}&&\nonumber\\
        &\discharge_{t+k} \leq \dischargerate \hspace{75pt}, \forall k \in \{0,...,N-1 \} &&\nonumber\\
        &SoC_{t+k} \leq \maxcharge \hspace{64pt} , \forall k \in \{0,...,N-1 \}&&\nonumber\\
        &\steer_{t+k} \leq \overline{\steerable} \cdot n_{t+k} \hspace{38pt} , \forall k \in \{0,...,N-1 \}&&\nonumber\\
        &\steer_{t+k} \geq \underline{\steerable} \cdot n_{t+k}  \hspace{38pt} , \forall k \in \{0,...,N-1 \}&&\nonumber\\
        &n_{t+k} \in \{0, 1\}  \hspace{62pt} , \forall k \in \{0,...,N-1 \}&&\nonumber
		\end{align}
	\STATE \textbf{Output:} $a^{N}_t$
	\end{algorithmic}
	\label{mpc}
\end{algorithm}
    The model-predictive controller (MPC) is used to define the control actions ($\discharge_t$, $\charge_t$, $\steer_t$) at each decision time-step $t$ by solving an optimization problem with $N$-step look-ahead. This controller receives as input the microgrid parameters and a forecast of the stochastic variables for the $N$ following time steps. The forecast for the consumption, is denoted by $\widehat{\nonFlexible}_{t}$, and is given by $\widehat{\nonFlexible}_{t} = (\widehat{\nonFlexible}_{t+k}, \forall k \in \{0,...,N-1 \})$. Accordingly, the forecast of the renewable production is denoted by $\widehat{\nonSteerable}_{t}$, and is given by $\widehat{\nonSteerable}_{t} = (\widehat{\nonSteerable}_{t+k}, \forall k \in \{0,...,N-1 \})$. 
    
    The optimization problem that is solved at each time-step is presented in Algorithm \ref{mpc}. The objective function aims at minimizing the curtailment, load shedding and fuel cost subject to the operational constraints defined by a mixed-integer linear model of the microgrid. The integer variables $n_{t+k}$ are used to ensure that when the generator is activated the generation level lies between its minimum stable generation level and its capacity. 
    
    The output of this controller is an open loop policy $a^{N}_t = ((\discharge_{t+k},\charge_{t+k},\steer_{t+k}), \forall k \in \{0,...,N-1 \})$ for the subsequent $N$ time-steps. At each control time-step $t$, only the first action from the sequence of computed actions is applied to the system $a_t= \left(\charge_{t}, \discharge_{t}, \steer_{t} \right)$. The quality of this controller depends on the number of look-ahead steps $N$, the accuracy of the forecasts and the quality of the model considered. 
\section{Case study}\label{sec: CaseStudy}

\subsection{System configuration}\label{sec: config}

The evaluation of the developed methodology is performed using empirical data measured by the off-grid micro-grid system of the village ``El Espino" (-19.188, -63.560), in Bolivia, installed in September 2015 and composed of photovoltaic (PV) panels, battery storage and a diesel generator. The system serves a community of 128 households, a hospital and a school, as well as the public lighting service. A comprehensive description of the system and of the data is available in previous work~\cite{Balderrama2019}. Aggregate electric load data is available as an indirect measurement, i.e. as the sum of direct measurements retrieved from the PV arrays, the diesel generator and the battery by means of smart meters. 

In this paper, we use the available measured data for the consumption and the PV production for the period 01/01/2016-31/07/2017. However, the existing components of the ``El Espino" microgrid i.e.~the installed equipment (battery, generator etc.), are rather oversized. The possibility to generate and store a lot of energy at once makes it easier to find an optimal policy. 

In an off grid-microgrid setting, the optimal size of the components depends heavily on the control policy applied. When the capacity of the installed components is large, a myopic policy can be as good as a look-ahead policy. On the other hand, a good policy that is able to anticipate changes and to act accordingly allows for the reduction of the components size and subsequently the installation cost.

The search for a good policy becomes much more relevant when the size of the components is constraining the operation of the microgrid. Therefore, in this paper we consider a reduced installation for which the applied control policy really impacts the cost of operation for the microgrid. The parameters used for the microgrid configuration in this paper are given in Table \ref{input_param}.

Additionally, the effect of different policies depend on the seasonality of solar irradiation and demand being observed. For instance, during the summer period (November through March in the case of Bolivia) there are high solar irradiation levels that can be used to charge fully the battery most of the days. During this period a myopic rule based strategy has very similar outcomes with a look-ahead strategy. However, during the winter period (April to October), when solar irradiation is limited and the battery may not be fully charged, a more elaborate strategy is necessary in order to guarantee low-cost security of supply in the microgrid.


\begin{table}
	\begin{center}
		\renewcommand\arraystretch{1}
		\caption{Input parameters.}
		\begin{tabular}[b]{l r r}
			\hline
			$\maxcharge$ & 120 & kWh \\
			$\chargerate$, $\dischargerate$ & 100 & kW \\
			\chargeEfficienty, \dischargeEfficienty & 75 \% \\ 
			$\OPprice{fuel}$ & 1 & \texteuro/kWh\\
			$\OPprice{curt}$ & 1.5 & \texteuro/kWh \\
			$\OPprice{shed}$ & 10 & \texteuro/kWh \\
			$\OPduration$ & 1 & h\\
			$\overline{\nonSteerable}$ & 120 & kW \\
			$\overline{\steerable}$ & 9 & kW \\
			$\underline{\steerable}$ & 0 & kW \\
			\hline
		\end{tabular}
		\label{input_param}
	\end{center}
\end{table}

	\subsection{Partial Observability}
	    As described in Section~\ref{sec: Simulator}, the process under consideration is non-stationary. The stochastic component of the transition kernel is known to be non-Markovian and the optimal decision requires knowledge of the next $l$ time steps. In supervised learning problems this issue is commonly addressed by state based networks \cite{taylor2018forecasting}. However, in this paper we take a similar approach as the one considered in the optimization based controller (Section \ref{sec: optcontroller}). We use the model $M_\psi$ as a 1-step forecaster. After a number of warm-up iterations $B$, we use the model to produce a forecast of the state in the $l$ following time-steps. This forecast is used to augment the actual state which is used to train the controller. A critical assumption of our approach is that the gradual changes to the system dynamics are sufficiently smooth for a single model $M_\psi$ to successfully track these changes.

	\subsection{Action Space and Meta-Actions}
	
%
%

	Due to the continuous and high dimensional nature of the state and the action spaces of the problem, reinforcement learning methods cannot be applied in their exact form. However, recent developments in the field of reinforcement learning have made possible the design of approximate optimal policies using function approximation.

	In our setting, function approximation alone does not suffice. The action space visited by the optimal controller from Section~\ref{sec: optcontroller} is constrained to a subspace of $\mathbb{R}^2$. Therefore, we elaborate on the design of a small and discrete set of actions $A'$ that maps to the original action space $A$.  This step is necessary for the use of policy-based algorithms, as the maximization problem defined in \eqref{eqn: optimalpolicy} is hard to solve.
	
	The meta-action $a'_{t}$ for each decision step $t$ is defined as: $$a'_{t} \in A'= \left\lbrace ``C",``D", ``G" \right\rbrace.$$ 
	
	Meta-action $``C"$ indicates the action to charge energy in the battery, when there is excessive renewable production ($\Delta P_t > 0 $). With meta-action $``D"$ we select to prioritize the discharge of the battery for covering the deficit of energy ($\Delta P_t < 0 $) in the microgrid. In case the battery does not suffice for covering this deficit, the generator will be activated. Alternatively, meta-action $``G"$ is used to prioritize the generator for supplying the the deficit of energy and the battery will be discharged only in the case that the maximum generating limit ($\overline{\steerable}$) is reached.
	

	In particular, at each decision step $t$ we provide as inputs to the dispatch Algorithm \ref{algoRBC}, the observed residual generation $\Delta P_t$ and the meta-action $y_t=a'_t$. The residual generation $\Delta P_t$ is computed after the realization of the stochastic variables ($\nonSteerable_t$, $\nonFlexible_{t}$) as $\Delta P_t = \nonSteerable_t - \nonFlexible_{t}$.

    Defining the action space in this way allows the use of the dispatch rule defined in Algorithm \ref{algoRBC} to obtain the control actions $a_t = (\charge_{t}, \discharge_{t}, \steer_{t})$. The discrete action space $A'$ simplifies the problem but restricts the class of possible policies, which sometimes harms the performance of the reinforcement learning methods.	We leave the problem of directly optimizing continuous actions as future work.
	


\subsection{Comparison with the benchmarks}

The algorithm is compared against the two benchmarks described in Section \ref{sec: Simulator} and the simple model free version of PPO \cite{Schulman2017}. An optimization controller with perfect knowledge and 1 period of look-ahead (``MPC-1") is considered in order to obtain a fair comparison to the proposed algorithm. An optimization controller with 24 periods of look-ahead and perfect knowledge (``MPC-24") is used to provide an upper bound on the performance of any control strategy. Additionally, a myopic rule-based controller, indicated in the results as ``heuristic", is used to provide a lower bound. We use PPO to denote the baseline algorithm which only performs model-free updates and \textsc{D-Dyna} to denote our method.

The label ``training step" on the x-axis refers to the number of times a new set of trajectories has been used for computing one or multiple gradient steps. For a fair comparison we fix the total number of samples available for the agent and compute the number of samples per training update accounting for the number of gradient step and the number of planning steps. Finally, results are averaged for 10 random seeds in order to account for stochasticity.

\subsection{Generalization}

One of the challenges of real world applications is the occurrence of changes in the transition dynamics. As described in Section \ref{sec: Simulator}, the dynamics of the microgrid are composed of a deterministic part and a stochastic part. The stochastic part is not controllable and therefore constitutes a source of progressive change. 

In this section we evaluate the ability of model based algorithm to adapt to gradual changes that occur in the state space. An algorithm that generalizes over unseen data distributions can provide a good initialization for fine-tuning the new controller.
The following protocol was used for training and evaluation of the proposed algorithms. We split the original dataset in a training set and a test set: the training set ranges from January 2016 to December 2016, while the test set ranges from January 2017 to July 2017. 

\begin{figure}[t]
	\includegraphics[width=0.51\textwidth]{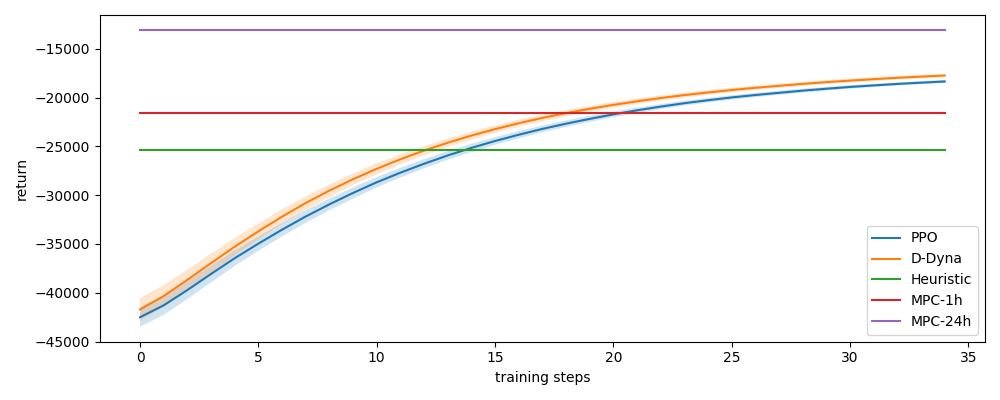}
	\centering
	\caption{Cumulative return on the test set.}
	\label{fig:rl-results}
\end{figure}

The results from the evaluation of the algorithms in the test set are presented in Figure~\ref{fig:rl-results}. We observe that the reinforcement learning methods are performing much better than the rule-based controller and the model based method is comparable to the upper bound set by the MPC-24. As illustrated in Figure~\ref{fig:rl-results}, introducing a model benefits generalization and both the baseline and the proposed algorithm are able to outperform the heuristic. We conjecture that using artificially generated states accelerates the learning process and provides a wider coverage of the state (exploration) and action space manifold, resulting in better generalization properties. 

Additionally, we observe that the proposed model based method is able to outperform the ``MPC-1" benchmark. We can argue that the obtained policy manages to resemble a look-ahead policy that takes optimal actions with respect to several steps ahead. This outcome is rather valuable because by using such a policy we can reduce the investment cost for equipment (e.g. battery capacity or diesel generators), without jeopardising the security of supply in the microgrid.

\subsection{Robustness}

In this section, we evaluate the performance of the proposed model based algorithm in sudden changes as defined in Section~\ref{sec: Simulator}. An example of such a change is the abrupt failure of the storage system, where the battery capacity is suddenly unavailable. 

We simulate this change in the following way. Let $x_{t}$ be the random discrete variable taking at each time-step the value 0 if the battery has failed and the value 1 if the battery is still operational. We assume that $x_{t}$ follows a Bernoulli probability distribution where $Pr(x_{t}=1)=p_{t}$, with $p_{t}$ following a linear decay in time and $p_{0}= 0.99$. If the battery fails, then the maximum storage capacity is considered to be reduced to zero ($\maxcharge=0 kWh$). After a failure, it is assumed that the battery equipment is fixed and the storage capacity is restored to its initial value in a period of $N=370$ hours. 

For this experiment, we have increased the size of the generator at a level that covers the entire demand. In this way, we want to evaluate the capability of the proposed model based method to switch from a regime where, the battery is mainly used when it is available, to only using the generator in the event of a battery failure.

\begin{figure}[t]
    	\includegraphics[width=0.51\textwidth]{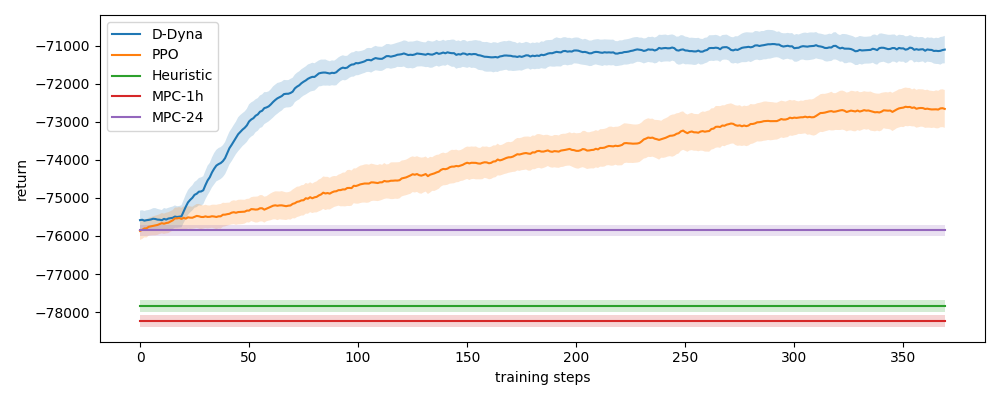}
    	\centering
    	\caption{Cumulative return when the battery is excluded.}
		\label{fig:change}
    \end{figure}

Under this scenario, we evaluate the benchmark controllers, the model-free method as well as the model based method. As we can see in Figure~\ref{fig:change}, all benchmarks perform poorly while the proposed algorithm is able to quickly adapt to the new drastically changing dynamics. The poor performance of the benchmark controllers is justified by the fact that there is no special equipment for the detection of the failure. The superiority of the proposed model based algorithm stems from its ability to detect the change since the model has been exposed to similar incidents during training.

\subsection{Transfer}

    In Reinforcement Learning, transfer learning is the ability of speeding up learning on new MDPs by reusing past experiences between similar MDPs. For real world applications, it would be desirable to obtain an algorithm that has the ability to learn off-line and adapt as the task changes. 

    A natural instance of such feature is to consider each month as a separate MDP and evaluate the ability to transfer knowledge across months. Note that each month has a different distribution of the stochastic component of the transition kernel.
    
	We set up the following experimental protocol. We use January 2016 to pre-train the algorithms. Then we initiate the training process for February and August 2016 using the pre-trained model. Intuitively transfer should be easier if the data distributions are close in time, and harder otherwise.
	
\begin{figure}
     	\centering
     	\begin{minipage}{0.5\textwidth}
     		\centering
     		\includegraphics[width=1.05\textwidth]{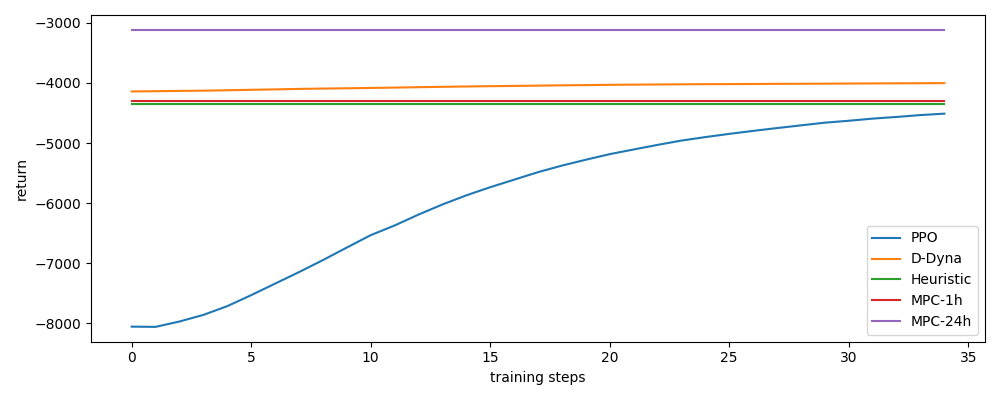}
     		\caption{Cumulative return on February.}
			\label{fig:transfer-feb}
     	\end{minipage} \hfill
     	\begin{minipage}{0.5\textwidth}
     		\centering
     		\includegraphics[width=1.05\textwidth]{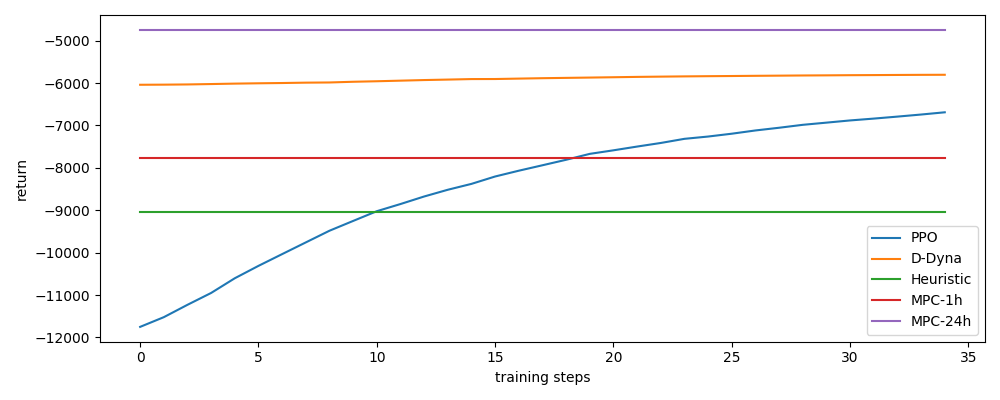}
     		\caption{Cumulative return on the August.}
			\label{fig:transfer-aug}
     	\end{minipage}
     
     \end{figure}

    The results of the described protocol are presented in Figures~\ref{fig:transfer-feb} and~\ref{fig:transfer-aug}. As we can see, transferring the model and the control allows for better performance than learning from scratch. As illustrated, the model based method can substantially speed up the learning process. The proposed method is shown to slightly outperform the ``heuristic" as well as the ``MPC-1h" benchmarks. However, in August the results are much better in that the model based method is approaching the performance of the ``MPC-24h" policy, while the rest of the benchmarks are falling behind. 
    
    As discussed in Section \ref{sec: config}, the effect of different policies depends on the period of the year. We can observe that the results in February are substantially different in comparison to August. There is a small discrepancy between the returns from the myopic and the optimization-based controller with perfect knowledge during February. On the other hand, during August the two policies show an increased difference in returns.

\section{Conclusions} \label{sec: conclusions}
    In this paper, a novel model based reinforcement learning algorithm is proposed for the lifelong control of a microgrid. First, an open-source reinforcement framework for the modeling of an off-grid microgrid for rural electrification is presented. The control problem of an isolated microgrid is casted as a Markov Decision Process (MDP). The proposed algorithm learns a model online using the collected experiences. This model is used to sample states during the evaluation step of the proximal policy optimization (PPO) algorithm.
    
    We compare the proposed algorithm to the standard benchmarks in the literature. Firstly, a rule-based control that takes decisions in a myopic manner based only on current information and secondly an optimization-based controller with look-ahead are considered for comparison purposes.
    
    We evaluate the generalization capabilities of the proposed algorithm by comparing its performance in out-of-sample data to the benchmarks. It is found that the use of the model to create artificial states leads to improved exploration and superior performance compared to the myopic rule-based controller and the MPC with one step look-ahead.
    
    We evaluate the robustness of the proposed algorithm when being subject to sudden changes in the transition dynamics, such as equipment failure. The results indicate that the model based method has the ability to adapt rapidly to severe changes in contrast to the benchmarks that are unable to detect changes and perform poorly to the subjected task.
    
    Finally, we evaluate the ability to transfer knowledge from one training session to the next. The results show large gains in computational time when initiating training on a new dataset with a pre-trained model.

	
	One important conclusion is that the proposed model based reinforcement learning method is able to adapt to changes both gradual and abrupt. Overall, the proposed method succeeds in tackling the key challenges encountered in the lifelong control of an off-grid microgrid for rural electrification. Future work should be directed to the design of a low dimensional continuous action space in order to be able to obtain results similar to the optimization-based controller.

\section{Acknowledgments}
This research is carried out in the framework of the Dynamically Evolving Long-Term Autonomy (DELTA) project.
DELTA is a European research project funded under the CHIST-ERA scheme (\url{http://www.chistera.eu/}). Anders Jonsson is partially supported by the grants TIN2015-67959 and PCIN-2017-082 of the Spanish Ministry of Science. The authors would like to thank Sergio Balderrama for the provision of measured data from the ''El Espino" microgrid in Bolivia and Alessandro Davide Ialongo for the fruitful discussion.


\bibliography{reference}
\bibliographystyle{src/icml2020}

\section*{Appendix}

\subsection*{Notation}
\subsection*{\textit{\textbf{Set and indices}}}

\begin{itemize}
	\item $t$, decision time step
	\item $k$, look-ahead step
	\item $\mathcal{A}$, action space
	\item $\mathcal{A'}$, meta-action space
	\item $\mathcal{S}$, state space
\end{itemize}  

\vspace*{0.25cm}
\subsection*{\textit{\textbf{Parameters}}}
\begin{itemize}
    \item $F_1$, $F_2$, fuel consumption parameters
    \item $N$, number of look-ahead periods
    \item $\widehat{\nonFlexible}$, load forecast (kW)
	\item $\chargerate$, $\dischargerate$, maximum charge and discharge rate (kW)
	\item $\overline{\nonSteerable}$, non steerable generation (kW)
	\item $\overline{\steerable}$, steerable generator capacity (kW)
	\item $\underline{\steerable}$, minimum steerable generation (kW)
	\item $\maxcharge$, $\mincharge$, maximum and minimum battery capacity (kWh)
	\item $\widehat{\nonSteerable}$, renewable generation forecast (kW)
	\item $\OPduration$, simulation and control period duration (h)
	\item $\chargeEfficienty$, $\dischargeEfficienty$, charge and discharge efficiency (\%) 
	\item $\OPprice{curt}$, curtailment price (\texteuro/kWh)
	\item $\OPprice{fuel}$, fuel price (\texteuro/kWh)
	\item $\OPprice{shed}$, load shedding price (\texteuro/kWh)
\end{itemize}
\subsection*{\textit{\textbf{Variables}}}
\begin{itemize}
	\item $a$, control actions vector
	\item $a'$, meta-actions vector
	\item $\nonFlexible$, non-flexible load (kW)
	\item $\OPcost{fuel}$, fuel cost (\texteuro)
	\item $\OPcost{curt}$, curtailment cost (\texteuro)
	\item $\OPcost{shed}$, lost load cost (\texteuro)
	\item $F_t$, fuel consumption ($l$)
	\item $k$, binary variable 
	\item $n_t$, number of cycles of the battery
	\item $\charge$, $\discharge$, charging and discharging power (kW)
	\item $\shed$, load shed (kW)
	\item $\curtail$, generation curtailed (kW)
	\item $\steer$, generation activated (kW)
	\item $\nonSteerable$, renewable generation (kW)
	\item $\charge$, charged energy of battery (kWh)
	\item $\discharge$, discharged energy of battery (kWh)
	\item $SoC$, state of charge of battery (kWh)
	\item $s$, control state vector 
	\item $\bar{s}$, stochastic state vector
	\item $\ubar{s}$, deterministic state vector
	\item $y_t$, discrete decision about the use of the equipment
	\item $\Delta P_t$, residual generation level (kW)
\end{itemize}

\subsection*{\textit{\textbf{Functions}}}
\begin{itemize}
	\item $P^{C}_{t}(\cdot)$, load probability distribution
	\item $P^{\nonSteerable}_{t}(\cdot)$, renewable generation probability distribution
	\item $s(\cdot)$, storage capacity as a function of the number of cycles
\end{itemize}
\end{document}